\documentclass{article}
\usepackage{ijcai20}

\usepackage{times}

\usepackage{soul}
\usepackage{url}
\usepackage[hidelinks]{hyperref}
\usepackage[utf8]{inputenc}
\usepackage[small]{caption}
\usepackage{graphicx}
\usepackage{amsmath,amsfonts}
\DeclareMathOperator{\sgn}{sgn}
\DeclareMathOperator*{\argmin}{arg\,min}

\usepackage{booktabs}
\usepackage{algorithm2e}

\title{
Quantum-Assisted Greedy Algorithms
}

\author{
	Ramin Ayanzadeh \and
	Milton Halem\and
	John Dorband\And
	Tim Finin\\
	\affiliations
		Department of Computer Science and Electrical Engineering\\
		University of Maryland, Baltimore County, Baltimore, MD 21250, United States\\
	\emails
		\{ayanzadeh, halem, dorband, {f}{i}nin\}@umbc.edu
}

\begin{document}

\maketitle

\begin{abstract}
We show how to leverage quantum annealers to better select candidates in greedy algorithms. Unlike conventional greedy algorithms that employ problem-specific heuristics for making locally optimal choices at each stage, we use quantum annealers that sample from the ground state(s) of a problem-dependent Ising Hamiltonians at cryogenic temperatures and use retrieved samples to estimate the probability distribution of problem variables. More specifically, we look at each spin of the Ising model as a random variable and contract all problem variables whose corresponding uncertainties are negligible. Our empirical results, on a D-Wave 2000Q quantum processor, revealed that the proposed quantum-assisted greedy algorithm (QAGA) can find notably better solutions (i.e., samples with lower energy value), compared to the state-of-the-art techniques in the realm of quantum annealing.
\end{abstract}

\section {Introduction}
Quantum artificial intelligence (QAI) is an emerging field that aims to apply quantum computers for addressing challenging problems in artificial intelligence (AI) \cite{lamata2017basic,biamonte2017quantum,dunjko2018machine} and use AI to advance quantum information processing \cite{king2019quantum,ayanzadeh2020reinforcement}.

Quantum annealing is a meta-heuristic that addresses combinatorial optimization problems (i.e., discrete optimization problems)—which are intractable in the realm of classical computing—by taking advantage of quantum mechanical fluctuations. Quantum annealers are a physical realization of the quantum annealing process that draws samples from the ground state(s) of a given Ising Hamiltonian at cryogenic temperatures (i.e., near-zero Kelvin) in near-constant time \cite{finnila1994quantum,kadowaki1998quantum,ohzeki2011quantum,nishimori2017exponential}. 

The quantum processing unit (QPU), by D-Wave Systems, is a programmable quantum annealer that samples from the ground state(s) of a given Ising Hamiltonian at cryogenic temperatures \cite{johnson2011quantum}. 
From a problem-solving point of view, the D-Wave quantum annealers receive coefficients of an Ising Hamiltonian as an executable quantum machine instruction (QMI), here $\mathbf{h}$ and ${J}$, and return the ground state of the following quadratic energy function:
\begin{equation}	
	\label{eqn:ising_energy}
	E_{\mathrm{Ising}}{(\mathbf{z})} = \sum_{i=1}^{N}{\mathbf{h}_i\mathbf{z}_i} + \sum_{i=1}^{N}{\sum_{j=i+1}^{N}{J_{ij}\mathbf{z}_i\mathbf{z}_j}},
\end{equation}
where ${N}$ denotes the number of quantum bits (qubits). In this representation, $\mathbf{z}_i \in \{-1, +1\}$. One can apply a linear transform to map Eq. \eqref {eqn:ising_energy} to its equivalent quadratic unconstraint binary optimization (QUBO) form, and vice versa, as follows:
\begin{equation}	
	\label{eqn:qubo_energy}
	E_{\mathrm{QUBO}}{(\mathbf{x})} = \sum_{i \leq j}^{N}{\mathbf{x}_iQ_{ij}\mathbf{x}_j},
\end{equation}
where $\mathbf{x} \in \{0, 1\}^N$ and $i,j \in \{1, 2, \cdots, N\}$. In this representation, ${Q}$ includes both linear biases and quadratic couplers—analogous to $\mathbf{h}$ and ${J}$ in Eq. \eqref{eqn:ising_energy}, respectively. 

Unlike conventional computing machines (namely classical and gate model quantum computers) that have a rich set of machine instructions, quantum annealers are single-instruction computing machines that can only minimize  Eq. \eqref {eqn:ising_energy} or \eqref {eqn:qubo_energy}.
One can reduce any problem of class NP to an NP-complete problem in polynomial-time \cite{garey2002computers,ayanzadeh2019sat}; thus, we can employ quantum annealers to address (significant) real-world problems that are intractable in the realm of classical computing.

As an illustration, we can employ the quantum annealers to address the original problem of binary compressive sensing (i.e., the $\ell_0$-norm sparse recovery) \cite{ayanzadeh2019quantum}. Similarly, we can represent the problem of non-negative binary matrix factorization as finding the global minimum of \eqref{eqn:qubo_energy} \cite{o2018nonnegative}. In the realm of quantum artificial intelligence and quantum machine learning, the D-Wave quantum annealer has been shown to address several problems including, but not limited to: planning \cite{rieffel2015case}, scheduling \cite{venturelli2015quantum,tran2016hybrid}, constraint satisfaction problems (CSP) \cite{bian2016mapping}, and training deep neural networks \cite{adachi2015application}. Moreover, the D-Wave quantum annealer has also demonstrated a capable performance in solving discrete optimization problems \cite{bian2014discrete}. In addition to the optimization aspect of the quantum annealers, one can employ the D-Wave QPU to sample from high-dimensional probability distributions, which has many applications in statistics, signal processing, artificial intelligence and machine learning \cite{biswas2017nasa,adachi2015application}.

We can form an Ising Hamiltonian whose ground state represents the optimum solution of any given problem; however, in practice, executing the corresponding QMI on a physical quantum annealer does not guarantee to achieve the global optimum \cite{ayanzadeh2020reinforcement}.
Current generations of the D-Wave quantum processors, as an example, have some technological barriers—including, but not limited to, sparse connectivity between qubits \cite{cai2014practical}, confined annealing schedule \cite{nishimori2017exponential}, coefficients’ range and precision limitations \cite{pudenz2015quantum,dorband2018extending}, noise and decoherence \cite{deng2013decoherence,gardas2018defects,gardas2018quantum}—that lower the quality of results (i.e., the energy value of the drawn samples is higher than the energy value of the ground state).

Recent studies have demonstrated that applying pre-processing techniques such as gauge transforms, can reduce the impact of analog errors \cite{pelofske2019optimizing}. Such preprocessing techniques, however, are trivial because they do not alter the landscape of the Ising Hamiltonian. 
In addition, performing post-processing heuristics, namely the multi-qubit correction (MQC), can result in finding notably better solutions (i.e., samples with lower energy value) \cite{dorband2018method}. Applying MQC, nevertheless, cannot guarantee that we can find the global minimum. It is worth highlighting that MQC performs well only on sparse problems and many real-world applications can have non-sparse representations.

Greedy algorithms are a problem-solving paradigm where we make locally optimal choices at each stage and expect that they yield a globally optimum solution. Although most greedy algorithms fail to achieve the global optimum, in many applications they are the best choice due to their efficiency \cite{devore1996some}. As an example, greedy algorithms are widely used in sparse recovery applications, at the cost of lower recovery accuracy (compared to convex optimization methods in compressive sensing) \cite{mousavi2019survey}. Note that greedy algorithms can find globally optimum solutions if the problem exhibits optimal substructure.

This study introduces a novel hybrid approach, called quantum-assisted greedy algorithms (QAGA), that leverages quantum annealers to better select candidates in each stage of the greedy algorithms. 
At each stage, QAGA employs a quantum annealer to provide samples from the ground state of the problem and use these retrieved samples to estimate the probability distribution of problem variables. 
After fixing variables with negligible uncertainties, QAGA proceeds to the next stage where the quantum annealer will solve a smaller problem with sparser couplings. 
Our experimental results, using a D-Wave 2000Q quantum processor, showed that QAGA can find samples with remarkably lower energy values, compared to the best-known enhancements in the realm of quantum annealing. 

\section{Method}
Let $\mathcal{H}$ be an Ising Hamiltonian as follows:
\begin{equation}
	\label{eqn:hamiltonian_problem}
	\mathcal{H} := {\sum_{i=1}^{N}{\mathbf{h}_i\mathbf{z}_i} + \sum_{i=1}^{N}{\sum_{j=i+1}^{N}{J_{ij}\mathbf{z}_i\mathbf{z}_j}}},
\end{equation}
where $\mathbf{h}$ and ${J}$ represent linear and quadratic coefficients, respectively. 
Also, let $\mathbf{z}^*$ represents the ground state of $\mathcal{H}$ as follows:
\begin{equation}
	\label{eqn:min_Hamiltonian}
	\mathbf{z}^* = \argmin_{\mathbf{z}}{\mathcal{H}}.
\end{equation}
In this study, we aim to find $\tilde{\mathbf{z}}$ such that:
\begin{equation}
	\label{eqn:ising_semi_optimality}
	\left| \mathcal{H}_{\mathbf{z}^*} - \mathcal{H}_{\tilde{\mathbf{z}}} \right| \to 0.
\end{equation}
In other words, our objective is to find a sample whose corresponding Ising energy value, shown in Eq. \eqref{eqn:ising_energy}, approaches the energy value of the ground state of the given Ising Hamiltonian.
QAGA starts with:
\[
	\tilde{\mathbf{z}}=\{\},
\]
and,
\[
	\mathcal{H}^0 = \mathcal{H}.
\]

At each iteration, QAGA uses a quantum annealer to draw ${n}$ samples from the ground state of $\mathcal{H}^t$. 
To solve problem \eqref{eqn:min_Hamiltonian} on a D-Wave quantum processor, as an illustration, one needs to embed $\mathcal{H}^t$ to the working graph of the D-Wave quantum annealers(i.e., the Chimera topology for the current generation), and normalize $\mathbf{h}$ and ${J}$ to $[-2,+2]$ and $[-1,+1]$, respectively \cite{ayanzadeh2020reinforcement}.

Let ${Z}$ denotes the set of all samples, drawn by a quantum annealer from the ground state of $\mathcal{H}^t$, as follows:
\[
	Z=\{\mathbf{z}^1, \mathbf{z}^2, \cdots, \mathbf{z}^n \},
\]
where $\mathbf{z}^j \in \{-1,+1\}^N$. 
All samples (i.e., $\mathbf{z}^j$ for $j=1,2, \dots, n$) contain a measurement for every qubit.
Hence, we can look at each problem variable ($\mathbf{z}_i$, for $i=1, 2, \cdots, N$) as a random variable with Bernoulli distribution that takes its value from $\{-1, +1\}$. 
Note that in this representation, the value -1 in \eqref{eqn:ising_energy} is analogous to 0 in \eqref{eqn:qubo_energy}; therefore, we can extend all analysis to QUBO form and employ QAGA in binary settings.

After retrieving the sample set ${Z}$, we estimate the uncertainty of every problem variable ($\mathbf{z}_i$) as follows:
\begin{equation}
	\label{eqn:qaga_var_uncertainty}
	u(\mathbf{z}_i) = 1-\frac{|\sum_{j=1}^{n}{\mathbf{z}_i^j}|}{n}.
\end{equation}
For any variable $\mathbf{z}_i$ of $\mathcal{H}^t$ that:
\begin{equation}
	\label{eqn:qaga_uncertainty_theta}
	u(\mathbf{z}_i) \leq \theta,
\end{equation}
where $\theta \in [0,0.5)$ specifies the threshold parameter, we fix the value of optimum solution as follows:
\begin{equation}
	\label{eqn:qaga_update_z*}
	\tilde{\mathbf{z}}_i = \sgn{ \left({ \sum_{j=1}^{n}{\mathbf{z}_i^j}} \right) }.
\end{equation}
Since $\theta < 0.5$, $\tilde{\mathbf{z}_i}$ is guaranteed to take its value from $\{-1,+1\}$.
After fixing the value of $\mathbf{z}_i$:
\begin{itemize}
	\item we remove $\mathbf{h}_i$ from $\mathcal{H}^{t+1}$;
	\item we add the value of $\tilde{\mathbf{z}}_i J_{ij}$ (or $\tilde{\mathbf{z}}_i  J_{ji}$) to $\mathbf{h}_j$, for $j=1,2, \dots, N$ and $i \neq j$;
	\item we remove the coupler $J_{ij}$ (or $J_{ji}$) from $\mathcal{H}^{t+1}$.
\end{itemize}
QAGA terminates when $\mathcal{H}^{t} \equiv \mathcal{H}^{t+1}$. 
If QAGA ended without fixing all problem variables, we apply the MQC method  \cite{dorband2018method} on the samples from the last stage of the QAGA and assign values for the remaining problem variables. 

Contracting variables with negligible uncertainties results in a new Ising Hamiltonian ($\mathcal{H}^{t+1}$) which is smaller and sparser, compared to $\mathcal{H}^{t}$. 
Hence, at each iteration of QAGA, the remaining Ising Hamiltonian becomes easier to be solved with physical quantum annealers.
Note that contracting variables reduces the number of qubits (${N}$) correspondingly. 
Algorithm \ref{algo:qaga} illustrates the QAGA process. 
Finally, we can apply a classical local search on $\tilde{\mathbf{z}}$ to increase the probability of finding the ground state.  

\begin{algorithm}[h]
	\DontPrintSemicolon 
	\KwIn{$\mathcal{H}, \theta$}
	\KwOut{$\tilde{\mathbf{z}}$}
	$\tilde{\mathbf{z}} \gets \{\}$\;
	$\mathcal{H}^{t} \gets \mathcal{H}$\;
	$\mathcal{H}^{t+1} \gets \{\}$\;
	\While{$\mathcal{H}^{t+1} \ne \mathcal{H}^{t}$} {
		$\mathcal{H}^{t+1} \gets \mathcal{H}^{t}$\;
		$Z\gets\{\mathbf{z}^1, \mathbf{z}^2, \cdots, \mathbf{z}^n\} = \argmin_{\mathbf{z}} \mathcal{H}^{t}$\;
		\For{$i\gets0$ \KwTo ${N}$}{
			\If{$u(\mathbf{z}_i) \leq \theta$}{
				$\tilde{\mathbf{z}}_i \gets \sgn{ \left({ \sum_{j=1}^{n}{\mathbf{z}_i^j}} \right) }$\;
				$\mathcal{H}_{\mathbf{h}_i}^{t+1} \gets \{\}$\;
				\For{$j\gets0$ \KwTo ${N}$}{
					\If{$J_{ij} \in \mathcal{H}^{t+1}$}{
						$\mathcal{H}_{\mathbf{h}_j}^{t+1} \gets \mathcal{H}_{\mathbf{h}_j}^{t+1} + J_{ij}\tilde{\mathbf{z}}_i$\;
					$\mathcal{H}_{J_{ij}}^{t+1}  \gets \{\}$\;
					}
					\If{$J_{ji} \in \mathcal{H}^{t+1}$}{
						$\mathcal{H}_{\mathbf{h}_j}^{t+1} \gets \mathcal{H}_{\mathbf{h}_j}^{t+1} + J_{ji}\tilde{\mathbf{z}}_i$\;
					$\mathcal{H}_{J_{ji}}^{t+1}  \gets \{\}$\;
					}
				}

			}
	  }
	}
\If{$|\tilde{\mathbf{z}}| < N$}{
	$\hat{\mathbf{z}} \gets \mathrm{MQC}(Z)$\;
	$\tilde{\mathbf{z}} \gets \hat{\mathbf{z}} \cup \tilde{\mathbf{z}}$\;
}
	\Return{$\tilde{\mathbf{z}}$}\;
	\caption{Quantum-assisted greedy algorithm (QAGA) for minimizing an Ising Hamiltonian}
	\label{algo:qaga}
\end{algorithm}

\section{Results}
This study proposes a hybrid approach, called quantum-assisted greedy algorithms (QAGA), for improving the performance of physical quantum annealers in finding the global minimum (i.e., the ground state) of Ising Hamiltonians, shown in \eqref{eqn:ising_energy}.
It is important to emphasize that, therefore, our objective was not to introduce a new greedy algorithm that can address combinatorial optimization problems nor to guarantee that the quantum annealers will find the global minimum of the given Ising Hamiltonian.

The current generation of the D-Wave quantum annealers (the Chimera architecture) includes more than 2,000 qubits and about 6,000 couplers, while the next generation (the Advantage architecture) will include more than 5,000 qubits and about 40,000 couplers \cite{boothby2019next}. Since coupling every qubit to every other qubit is infeasible in practice, the D-Wave QPUs have a sparse connectivity architecture. One can entangle multiple physical qubits to form a virtual qubit with higher connectivity at the expense of a significant reduction in the number of virtual qubits relative to the actual number of available physical qubits. 
As an example, 2048 physical qubits on the Chimera topology are equivalent to 64 fully-connected virtual qubits.

Benchmarking optimization techniques with computationally hard problems that we know the optimum solution is a common practice.
However, recent studies have demonstrated that applying MQC \cite{dorband2018method} on the drawn samples by the D-Wave quantum processors always finds the optimum solution of such problems \cite{dorband2019applying}.
Note that the current generation of the D-Wave quantum annealers is limited to represent a clique of size at most 63 and MQC may fail to find the global optimum of benchmark problems when we notably increase the number of variables, which is beyond the capacity of the current quantum annealers.
It is worth highlighting that MQC performs well only on sparse problems—when we reduce the sparsity of the problem, i.e., increasing the number of non-zero quadratic coefficients in Eq.  \eqref{eqn:ising_energy}—performance of the MQC approaches to a local search heuristic like single-qubit correction (SQC).
 
Generating random problems results in hard problems for benchmarking quantum annealers \cite{king2019quantum,dorband2018extending,dorband2018method}.
Hence, as a proof-of-concept, we use randomly generated Ising Hamiltonians for evaluating the performance of the proposed method in this paper (QAGA) and compare it with MQC which is the best-known post-quantum error correction for quantum annealers. 

\subsection{Experiment A}
For every problem in this experiment, we generated a random graph of size 50 with a specified sparsity rate ($s \in \{0.05, 0.25, 0.5, 0.75, 1.0\}$). More specifically,  we randomly selected edges from a complete graph with $N=50$ nodes where the sparsity rate ${s}$ denoted the probability of selecting edges. Afterward, we set values of the corresponding biases and couplers randomly.
We generated three different types of benchmark problems: (a) we picked values of biases and couplers randomly from $\{-1,+1\}$ (binary coefficients); (b) we used random numbers in $[-1,+1]$ with uniform distribution to assign values of $\mathbf{h}$ and ${J}$ (uniform coefficients); and (c) coefficients’ values were drawn randomly from a standard normal distribution (normal coefficients).

Our objective in this study was to improve the quality of results (i.e., finding samples with lower energy value), attained by the D-Wave quantum annealers. Hence, as our ground-truth, we compared the performance of QAGA with two different settings. 
Since applying spin-reversal-transforms (a.k.a. gauge transforms) has demonstrated to address the impact of analog errors of the physical quantum annealers \cite{pelofske2019optimizing}, as our first ground-truth method, we compared QAGA with quantum annealing with 10 spin-reversal transforms (here, denoted by QA). 
We also enabled the inter-sample delay to reduce the sample-to-sample correlations in successive reads/measurements, albeit longer execution time. 
As our second ground-truth method, we applied MQC on samples from QA—quantum annealing with 10 spin-reversal-transforms and enabled inter-sample delay—denoted by MQC. 

For each QMI, we requested 1,000 samples for all methods.
The uncertainty threshold in QAGA was $\theta=0.0$—i.e., all spins must have the same value so QAGA can fix them for the next stage.
We obtained this threshold empirically via evaluating the performance of QAGA on a small problem set.
In the next experiment, we have studied the impact of $\theta$ on the performance of QAGA. 

Since randomly generated problems are not compatible with the working graph of the current D-Wave quantum processors, we used the minor-embedding heuristic \cite{cai2014practical} for embedding the arbitrary random graphs to the Chimera topology of the D-Wave 2000Q quantum processors. 
In addition, QAGA iteratively contracts variables whose uncertainties are negligible; thus, the Ising Hamiltonian in QAGA has a dynamic structure, in terms of the number of qubits and their connectivity.
In this sense, at each stage, QAGA needs to embed the remaining Ising Hamiltonian to an executable QMI on the target quantum annealer. 

\begin{figure*} [h]
	\centering
	\includegraphics[scale=0.75]{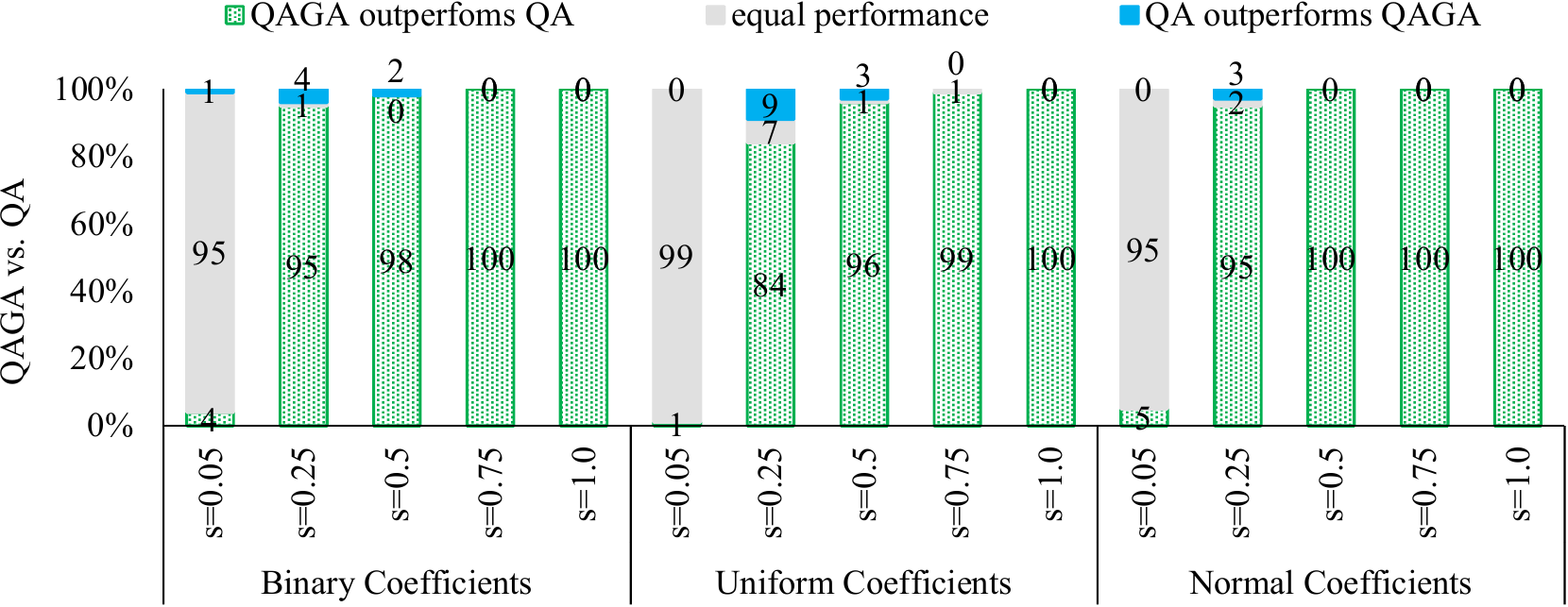}
	\caption{Performance comparison between QAGA and QA (with spin-reversal-transforms and inter-sample delays) in solving 100 random benchmark problems with $N=50$ spin variables and different sparsity rates ${s}$}
	\label{fig:qaga_vs_qa}
\end{figure*}

Figure \ref{fig:qaga_vs_qa} illustrates the performance comparisons between QAGA and QA in minimizing Ising Hamiltonians with binary, uniform and normal coefficients. 
For each case, we generated 100 random problems with the corresponding (uniform or normal) distributions. 
In this experiment, all randomly generated Ising Hamiltonians had 50 spin variables ($N=50$). 
For each sparsity rate, the corresponding column illustrates the number of times that QA has found a better sample (compared to QAGA), number of times that QA and QAGA demonstrated same performance (i.e., best samples from both methods had identical Ising energies), and number of times that QAGA has outperformed QA. 

\begin{figure*}[h]
	\centering
	\includegraphics[scale=0.75]{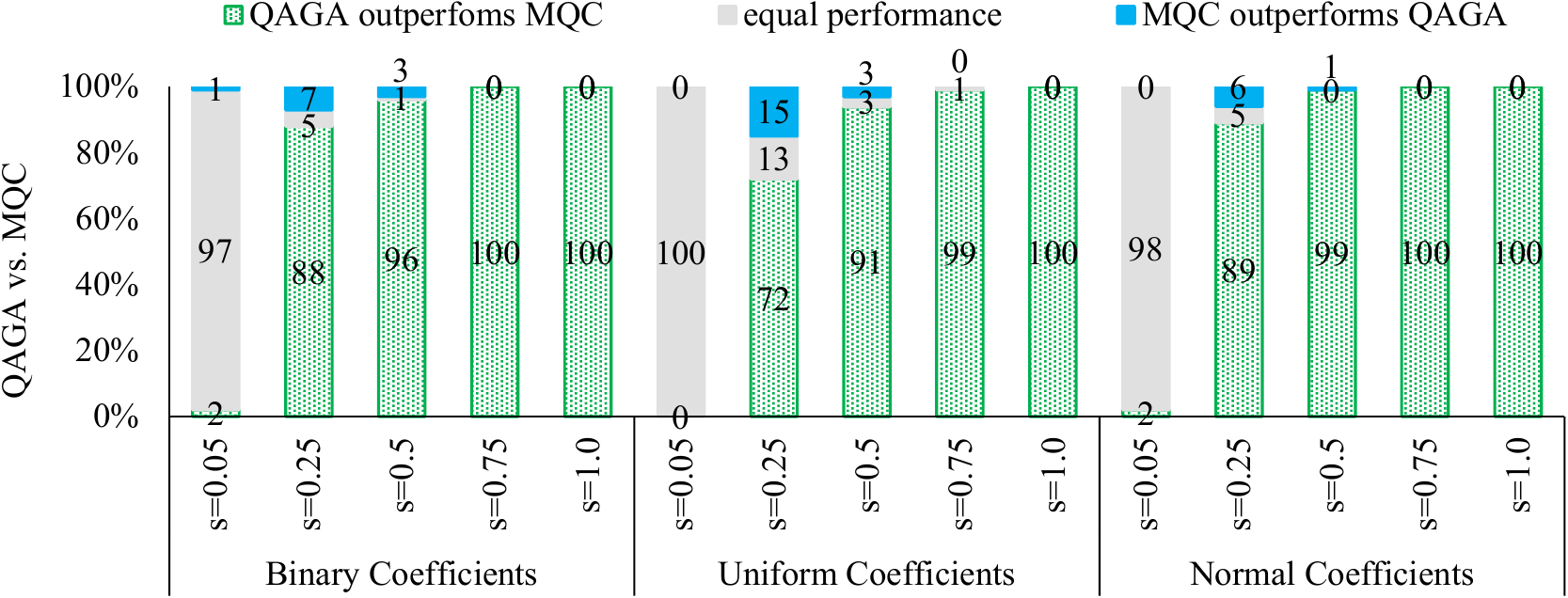}
	\caption{Performance comparison between QAGA and MQC in solving 100 random benchmark problems with $N=50$ spin variables and different sparsity rates ${s}$}
	\label{fig:qaga_vs_mqc}
\end{figure*}

In the same manner, Fig. \ref{fig:qaga_vs_mqc} illustrates the performance comparisons between QAGA and MQC in minimizing the same 100 randomly generated Ising Hamiltonians with binary, uniform and normal coefficients. 
Note that all arrangements in this experiment (e.g., number of variables, sparsity rate, number of spin-reversal-transforms, etc.) were identical to the previous experiment.
Similar to Fig. \ref{fig:qaga_vs_qa}, each column in Fig. \ref{fig:qaga_vs_mqc} represents the number of times that MQC has found a sample with lower energy, MQC and QAGA had similar performance, and QAGA resulted in a sample with lower Ising energy value. 

\subsection{Experiment B}
In this experiment, we aim to study the impact of the threshold parameter on the performance of QAGA. To this end, we measured the average number of stages (iterations) that QAGA takes to be converged. 
Table \ref{tbl:qaga_num_iterations} illustrates the average number of iterations that QAGA takes to solve 100 random benchmark problems with $N=50$ variables and normal coefficients.
Since the randomly generated Ising Hamiltonians had arbitrary graph structures, we used the minor-embedding heuristic \cite{cai2014practical} for mapping them to executable QMIs on the D-Wave quantum annealers. 
 
\begin{table}[h]
	\caption{Average number of iterations for QAGA in solving 100 random benchmark problems with different thresholds }
	\centering
	{
	\smallskip\begin{tabular}{c|ccccc}
			&	\multicolumn{5}{c}{sparsity rate (${s}$)}\\ 
		$\theta$	&	0.05	&	0.25	&	0.50	&	0.75	&	1.00\\ \cline{2-6}
		0.25		&	3.25	&	2.80	&	3.15	&	3.40	&	3.30 \\
		0.15		&	3.60	&	3.40	&	3.35	&	3.45	&	3.65 \\
		0.05		&	4.10	&	4.45	&	4.30	&	3.20	&	4.05 \\
		0.00		&	5.50	&	6.20	&	2.10	&	2.05	&	2.30 \\
\end{tabular}
	}
	\label{tbl:qaga_num_iterations}
\end{table}
Table \ref{tbl:qaga_num_iterations} reveals that for Chimera like problems (i.e., sparse problems where $s=0,05$ or $0.25$), when $\theta \to 0$, QAGA takes more iterations (i.e., slower convergence). 
On the other side, for dense Ising Hamiltonians, where $s \to 1$ (i.e., clique like problems), when $\theta \to 0$, QAGA converges quickly—the maximum number of iterations appears on $\theta \sim 0.1$. 

\section{Conclusion}
Quantum annealing is a meta-heuristic for addressing discrete optimization problems in near-constant time; however, owing to some technological barriers, physical quantum annealers (like the D-Wave quantum processors) generally result in excited states (i.e., close to global minimum), rather than converging to the ground state (i.e., global minimum).
In other words, quantum annealers can find very high-quality solutions for combinatorial optimization problems in a fraction of a second; nevertheless, they generally fail to get to the ground state of the given Ising Hamiltonians.

In this study, we introduced a novel hybrid approach, called quantum-assisted greedy algorithms (QAGA), that employs the quantum annealers for making globally optimum choices at each stage of a greedy algorithm.
To this end, we looked at the quantum annealers as a physical process that naturally draws samples from the ground state of Ising Hamiltonians (i.e., a problem-dependent Boltzmann distribution) at cryogenic temperatures.
From a problem-solving point-of-view, conjugating quantum annealers and greedy algorithms address drawbacks of both methods and results in remarkably better solutions (i.e., samples with lower energy value), at the cost of executing multiple QMIs for one problem. 

Our empirical results, using a D-Wave 2000Q quantum processor, on several randomly generated benchmark problems demonstrated that QAGA finds samples with remarkably lower energies, compared to the best-known techniques in the realm of quantum annealing—namely spin-reversal-transforms (or gauge transform) and multi-qubit correction (MQC).
For sparse problems (i.e., the structure of the problem is close to the Chimera architecture), the performance of the QAGA approaches to MQC.
When the sparsity decreases, however, QAGA shows supremacy in terms of finding samples with lower energy.
In other words, QAGA implicitly addresses the embedding-related issues, specifically the broken-chains.

\subsubsection*{Acknowledgements}
This research has been supported by NASA grant (\#NNH16ZDA001N-AIST 16-0091), NIH-NIGMS Initiative for Maximizing Student Development Grant (2 R25-GM55036), and the Google Lime scholarship. We would like to thank the D-Wave Systems management team, namely Rene Copeland, for granting access to the D-Wave 2000Q quantum annealer.

\bibliographystyle{named}
\bibliography{biblio}

\end{document}